\date{}
\documentclass[showpacs]{revtex4}
\usepackage{graphicx,psfrag,amsmath,amssymb,amsfonts,latexsym,color,dcolumn}

\begin{document}

\title{Spin bath interactions effects on the geometric phase}

\author{Paula I. Villar \footnote{Electronic address: paula@df.uba.ar}}
\affiliation{Departamento de F\'\i sica {\it Juan Jos\'e Giambiagi}, FCEN, UBA,
 Ciudad Universitaria, Pabell\' on I, 1428 Buenos Aires, Argentina.\\
Computer Applications on Science and Engineering Department, Barcelona
Supercomputing Center (BSC), 29, Jordi Girona 08034 Barcelona Spain}

\date{today}

\begin{abstract}
We calculate the geometric phase of a spin-1/2 particle coupled to
an external environment comprising N spin-1/2 particle in the framework
of open quantum systems. We analyze the decoherence factor and the
deviation of the geometric phase under a nonunitary evolution
from the one gained under an unitary one. We show the dependence
upon the system's and bath's parameter and analyze the range of
validity of the perturbative approximation. Finally, we discuss the
implications of our results.
\end{abstract}


\maketitle

\newcommand{\beq}{\begin{equation}}
\newcommand{\eeq}{\end{equation}}
\newcommand{\beqa}{\begin{eqnarray}}
\newcommand{\eeqa}{\end{eqnarray}}
\newcommand{\beqas}{\begin{eqnarray*}}
\newcommand{\eeqas}{\end{eqnarray*}}

\section{Introduction}

A quantum system undergoing a cyclic evolution behaves
quite differently from its classical analogue: the state vector
retains the history of its evolution in the form of a
geometric quantum phase.  Berry \cite{Berry}
demonstrated that closed quantum systems could
acquire phases that are geometric in nature. He showed that,
besides the usual dynamical phase, an additional phase
related to the geometry of the space state is generated during an
adiabatic evolution. A phase only dependent upon the
area covered by the motion of the system.

A renewed interest in geometric phases
(GP) has recently blossomed by the proposal of using GP
for quantum computation.
The use of fault tolerant quantum gates in quantum information
requires the study of the geometric phase in open
quantum systems \cite{LomMazziVillar, PRA, Sanders, Carollo, chiara, Whitney, Yi}.
 This has been motivated by the fact
that all realistic quantum systems are coupled to
their surrounding environments. No matter how
weak the coupling that prevents the system from being isolated, the
evolution of an open quantum system is  plagued by
nonunitary features like decoherence and dissipation. Decoherence,
in particular, is a quantum effect whereby the system looses its ability
to exhibit coherent behaviour and appears as soon as
the two interfering partial waves shift the environment into
states orthogonal to each other \cite{Visibility}.
Nowadays, decoherence stands as a serious obstacle in quantum
information processing.

Starting with the seminal paper of Zurek \cite{Zurek},
several authors analyzed the decoherence due to a collection
of independent spins.
In this paper we discuss a two-level system coupled
to a one-dimensional array of spin-1/2 particles.
This model evolves reigned by a Hamiltonian 
which encompasses Ising and Universality 
in  one simple approach. In some trivial
cases, we can solve  the problem exactly.
Most of the works done so far are based on the so called
``central spin model'', where the two level system
is coupled isotropically to all spins of the bath, which
implies a enormously simplification of the derivation.
To ensure decoherence, the central spin (the system)
is coupled to the transverse component of the bath.

Working towards having a realistic description of geometric
phases, we have introduced field decoherence by coupling
the central two level system to an external environment.
In this paper we investigate the geometric phase
of a spin-1/2 particle interacting with a spin environment.
The model is the most simple case
of an Ising chain and a further consideration
to the Zurek model, since we are considering the self energy
of the environment spin-1/2 particles. We calculate and analyze
the decoherence factor induced by the coupling to an
external bath. Within the general framework of
GPs in open quantum systems, we estimate the corresponding
GP for the spin-1/2 particle in the weak coupling regime.
We further compare this result to the general GP obtained
numerically and analyze its dependence upon the bath and
system parameters.

This paper is organized as follows. In Sec. II, we present the
model and analyze the decoherence factor due to the interaction
with the environment. In Sec. III, we
estimate the geometric phase for the central spin and
thoroughly study its behaviour by performing numerical
and analytical calculations.
Finally, we conclude our results in Sec. IV.

\section{Model}

In this section we examine the decoherence induced by
disordered interacting spin baths at finite temperature.
Our choice of the bath is the most simple case for an Ising
chain so as to facilitate an analytical study of the
decoherence and the geometric phase for the model. A further
study of the geometric phase for an Ising chain with random
spin-spin interactions will be studied elsewhere.
Our system is comprised by a central spin $\sigma_c$
coupled to a bath of N Ising spins
through the interaction Hamiltonian
\beq
H_{\rm I}=\sigma_c^{z}\otimes\sum_{i=1}^{N}\lambda_i \sigma_i^{z}.
\eeq
The coupling between the central spin and the bath is characterized
by the coupling constant $\lambda$. The presence of a
non interacting spin chain will be modeled
by a bath Hamiltonian such as
\beq
H_{\rm B}=\sum_{i=1}^{N} \omega_i \sigma_i^{x},
\eeq
where $\sigma_i^{x}$ denotes the Pauli operators of the bath
spins.
% For the moment we are neglecting the self-Hamiltonian
% of the system as, in this case, it is a
% reasonable aproximation \cite{Zurek}.
The self Hamiltonian of the central spin is the common one
$H_s= \hbar/2 \Omega \sigma_c^z$.
This model results a further approach to the model studied
in \cite{Zurek} since we are considering not only the interaction
Hamiltonian but also the self-Hamiltonians
of the N spins of the bath. In addition, we must note
that $[H_B,H_{\rm I}] \neq 0$.

We start with a total separable state of the form such as
\beq
\mid \Psi(0) \rangle = (\alpha \mid 0 \rangle + \beta \mid
1 \rangle) \times \mid \chi_0 \rangle,
\eeq
where $\mid 0 \rangle$ and $\mid 1 \rangle$
are the state eigenstates in the $\sigma^{z}$ basis and
$\chi_0$ is the initial state of the bath. At a later
time, the total state of the system is
\beq
\mid \Psi({\rm t}) \rangle = e^{-i H t} \mid \Psi(0) \rangle,
\eeq

where $H$ is the evolution total Hamiltonian. The main object in the study of decoherence is the reduced
density matrix which is obtained after tracing out the
environment's degrees of freedom. In our case, the
reduced density matrix may be written as
\begin{eqnarray}
\rho_r(t) = \bigg(
\begin{array}{c c}
\mid \alpha \mid ^2 & \alpha \beta^* F(t) \\
\alpha^* \beta F(t)^* & \mid \beta \mid ^2 \\
\end{array}
\bigg),
\end{eqnarray}
where $F(t)= {\rm Tr}_B [ e^{-i(H_B + V) t} \rho_B(0)
e^{i(H_B - V)t}]$. We have named $V=\sum_{i=1}^N
\lambda_i \sigma_i^{z}$ after the interaction Hamiltonian
has operated on the central spin.

We can choose a pure state for the initial state of the
bath as $\vert \chi_0 \rangle=\prod_{i=1}^N (\alpha_i
\vert 0_i \rangle + \beta_i \vert 1_i \rangle)$. Since
$\rho_B(0)= \vert \chi_0 \rangle \langle \chi_0 \vert$,
the decoherence factor $F(t)$ yields
\beq
F(t)= \prod_{i=1}^N \bigg( 1- \frac{2 \lambda_i^2}
{ \omega_i^2 + \lambda_i^2} \sin^2 (\sqrt{\omega_i^2
+ \lambda_i^2} t)\bigg).
\eeq

We can see that this decoherence factor does not depend
on the external temperature. It is also possible
to check that $F(0)=1$. As it was done in \cite{Zurek},
we can evaluate the mean value of the decoherence factor.
 This expression is similar
to that obtained in \cite{Chitra} but for a different
initial state of the environment. As in that case,
 $\langle F(t) \rangle_{T \rightarrow \infty}
 \rightarrow 0$, when $T \rightarrow \infty$. If we
estimate the average dispersion as
$\Delta^2 = \langle F(t)^2 \rangle - \langle F(t) \rangle ^2=
\sum_{i=1}^N p_i$, with $p_i=(1- \frac{\lambda_i^2}
{4 (\omega_i^2 + \lambda_i^2)})$. Under the assumption that
all $p_i$ are approximately equal, the average fluctuations
from zero are
\beq
\Delta \sim \frac{1}{\sqrt{N}}.
\eeq
Therefore, large environments can effectively induced
decoherence on the central spin system.

\section{Geometric phase environmentally corrected}

In this section, we shall compute the geometric phase
for the central spin and analyze its deviation from
the unitary geometric phase for a spin-1/2 particle.

A proper generalization of the geometric phase for
unitary evolution to that for non unitary evolution
is crucial for practical implementations of geometric
quantum computation. In \cite{Tong}, a quantum kinematic
approach was proposed and the geometric phase 
(GP) for a mixed state
under nonunitary evolution has been defined
by Tong {\it et.al.} as
\begin{eqnarray} \Phi & = &
{\rm arg}\{\sum_k \sqrt{ \varepsilon_k (0) \varepsilon_k (\tau)}
\langle\Psi_k(0)|\Psi_k(\tau)\rangle \times e^{-\int_0^{\tau} dt \langle\Psi_k|
\frac{\partial}{\partial t}| {\Psi_k}\rangle}\}, \label{fasegeo}
\end{eqnarray}
where $\varepsilon_k(t)$ are the eigenvalues and
 $|\Psi_k\rangle$ the eigenstates of the reduced density matrix
$\rho_{\rm r}$ (obtained after tracing over the reservoir degrees
 of freedom). In the last definition, $\tau$ denotes a time
after the total system completes
a cyclic evolution when it is isolated from the environment.
Taking the effect of the environment into account, the system no
longer undergoes a cyclic evolution. However, we will consider a
quasi cyclic path ${\cal P}:t ~\epsilon~[0,\tau]$ with
$\tau=2 \pi/\Omega$ ($\Omega$ the system's frequency).
When the system is open, the original GP, i.e. the one that
would have been obtained if the system had been closed $\Phi^U$, is
modified. This means, in a general case, the phase is
$ \Phi=\Phi^U+ \delta \Phi$,
where $\delta \Phi$ depends on the kind of environment coupled to
the main system \cite{PRA,barnerjee,zanardi}. For a spin-1/2 particle
in $SU(2)$, the unitary GP is known to be $\Phi^U= \pi(1+\cos(\theta_0))$.
It is worth noticing that the proposed GP is gauge invariant and leads
to the well known results when the evolution is unitary.

To this end, we shall start by choosing
a pure state of the form:
 \begin{equation}
|\Psi (0) \rangle=\cos(\theta_0/2) |0 \rangle
+ \sin(\theta_0/2) |1 \rangle .
\nonumber\end{equation}
This is the same to have assumed $\alpha=\cos(\theta_0/2)$
and $\beta=\sin(\theta_0/2)$ in our above deduction of
the decoherence factor. Then, for times $t>0$,
the state of the system is

\begin{equation}
\label{Phi+}
|\Psi_{+} (t) \rangle = e^{-i \Omega t} \cos(\theta_+(t))
|0\rangle + \sin(\theta_+(t)) |1\rangle,
\end{equation}
where we have defined $\tan(\theta_+(t))
= \tan(\theta_0/2)F^{-1}(t)$.
It is easy to check that, when $F(t)=1$, we re-obtain the results
for the unitary case. 
From Eq.(\ref{Phi+}), one can easily write the reduced
density matrix $\rho_r(t)$ and estimate its
eigenvalues and eigenvectors, as was done in Ref.\cite{PRA}.
In order to estimate the geometric phase, we only need the
eigenvector $|\Psi_{+}(t) \rangle$ since $\varepsilon_{-}(0)=0$,
 and, hence, the only contribution to the phase comes from that
 eigenvector and its corresponding eigenvalue (see
 Eq.(\ref{fasegeo})). In the case we are considering here, 
and after performing
the time derivatives of Eq. (\ref{fasegeo}), we can note
that the eigenvalues 
 and eigenvectors are real functions. Consequently,
the geometric phase is
\begin{equation}
 \Phi = \Omega \int_0^{\tau} \cos^2(\theta_+(t)).
\label{fasenu}
\end{equation}

As we can deduce from the relation between $\theta_+(t)$
and the decoherence factor, Eq.(\ref{fasenu}) is not
easily computed. In Fig.1, we show the
behaviour of the GP as a function of the initial
position on the Bloch Sphere ($\theta_0$) and the
value of the coupling constant ($\lambda$) for
an external environment
 comprising 10 spin-1/2 particles.

\begin{figure}[!ht]
 \centering
\includegraphics[width=10cm]{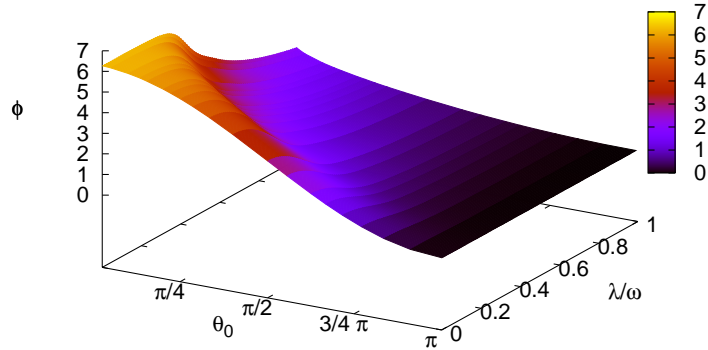}
\caption{Exact GP as a function of the coupling constant
 and the initial angle in the Bloch sphere. For simplicity,
we have assumed that all the bath spins have
 the same frequency and coupling constant and $N=10$.}
\label{figure1}
\end{figure}

In order to achieve an analytical result, we shall assume
all the bath spins to have the same coupling constant and
frequency. Then, the decoherence factor becomes
$F(t)= \prod_{i=1}^N f_i(t)= f(t)^N$ and we can forget
about the product function. Besides, we shall perform
a perturbative expansion in powers of the coupling
constant $\lambda$. In such a case,
\begin{eqnarray}
 \Phi &=& \pi (1+\cos(\theta_0)) + N \bigg(\frac{\lambda}{\omega}\bigg)^2
\sin^2(\theta_0)\bigg(\pi - \frac{ \Omega}{4 \omega}
 \sin(4 \pi \omega/ \Omega)\bigg). \nonumber \\
\label{faseper}
\end{eqnarray}

Notice that the lowest correction in $\lambda/\omega$
of Eq.(\ref{faseper}) is quadratic, which means that in
case of low decoherence we recover the unitary GP up
to the first order in $\lambda/\omega$. This reflects
the resilient of the phase against the environment.
If the decoherence rate is sufficiently small, the
probability amplitude of staying in the same eigenstate
is almost stationary. This amounts to effectively
project the state back to the original eigenstate, whenever
it tends to be driven away by the environment. Therefore,
the state trajectory on the projective Hilbert space
tends to be unaffected by the decoherence (up to first
order), thereby leaving the area enclosed by the path,
and hence the geometric phase, unchanged. Similarly,
this robustness against decoherence can be observed
in different two-level systems \cite{PRA, Carollo, chiara}.

\begin{figure}[!ht]
\centering
\includegraphics[width=10cm]{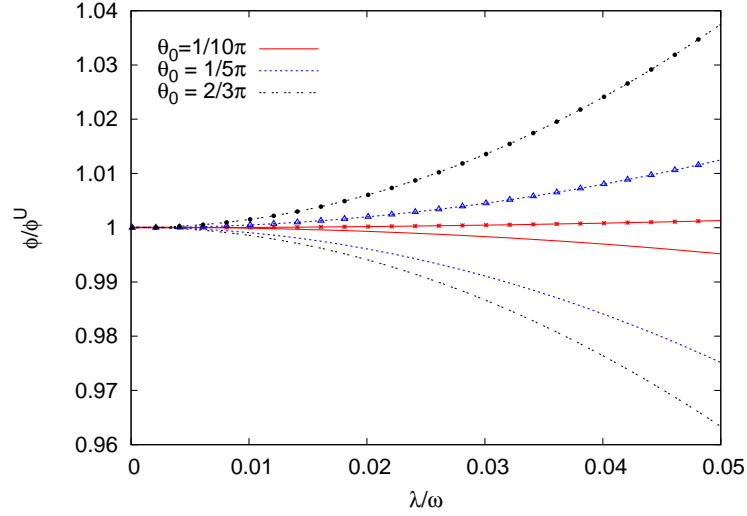}
\caption{Comparison between the exact GP and the perturbative
approximation for different initial positions on the Bloch
sphere and $N=10$ spin environments. The lines (whether
solid or dashed) represent the exact GP while the lines with
different shaped points are the corresponding perturbative
 perturbative calculations.}
\label{faseperN10}
\end{figure}

\begin{figure}[!ht]
\centering
\includegraphics[width=10cm]{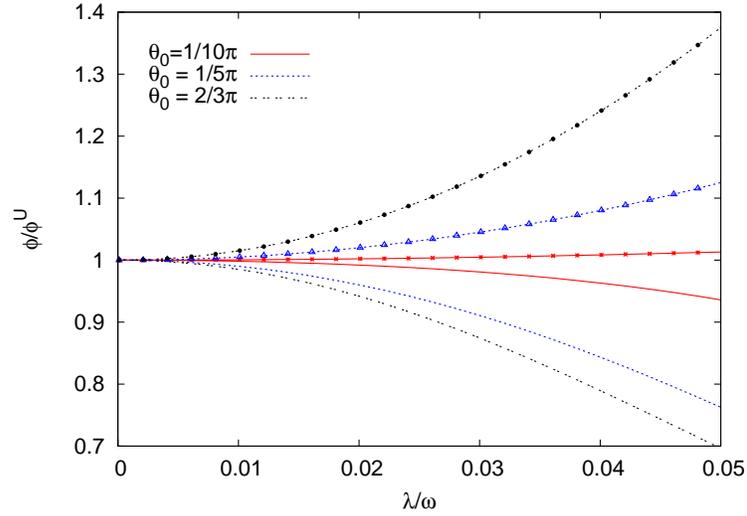}
\caption{Comparison between the exact GP and the perturbative
approximation for different initial positions on the Bloch
sphere and $N=100$ spin environments. The lines (whether
solid or dashed) represent the exact GP while the lines with
different shaped points are the corresponding perturbative
 perturbative calculations.}
\label{faseperN}
\end{figure}

In Figs.\ref{faseperN10} and \ref{faseperN}, we show the
comparison between the exact calculation (Eq.(\ref{fasenu}))
 of the GP and the perturbative approximation (Eq.(\ref{faseper}))
 for different values of the
coupling constant. In Fig.\ref{faseperN10}, we plot
the nonunitary GP for a case of a small environment consisting
of $N=10$ spins. The lined curves, whether solid or dashed, correspond
to exact calculations of the GP, for different initial position on the
Bloch sphere. The perturbative estimations are plotted with lines and
points in each case. Therein, it is possible to see that for a
small initial angle, i.e. small value of $\theta_0$, there is
almost no significant difference between both approaches.
However, for an initial state of bigger value of $\theta_0$,
the environmentally-induced correction to the unitary GP
becomes important which is shown in Fig.\ref{faseperN},
for an environment of $N=100$ spins. This difference
 among the approaches becomes more relevant 
as the number of spin environments increases and can be noted
 by comparing the scales of the y-axis in each figure.
This can easily be understood by looking at Eq.(\ref{faseper}).
Therein, one can identify a new ``effective'' coupling constant
$\lambda_{\rm eff}^2= N (\lambda/\omega)^2$, which means
that one must consider the product of
both quantities and that the perturbative approximation
will hold for a compromise between them. That is to say that
for larger environments (bigger values of $N$), the 
perturbative estimation will remain valid for smaller values
of the coupling constant. However, this estimation remains
valid for a bigger range of the coupling constant if we 
reduce the number of spins comprising the environment.

\begin{figure}[!ht]
\centering
\includegraphics[width=10cm]{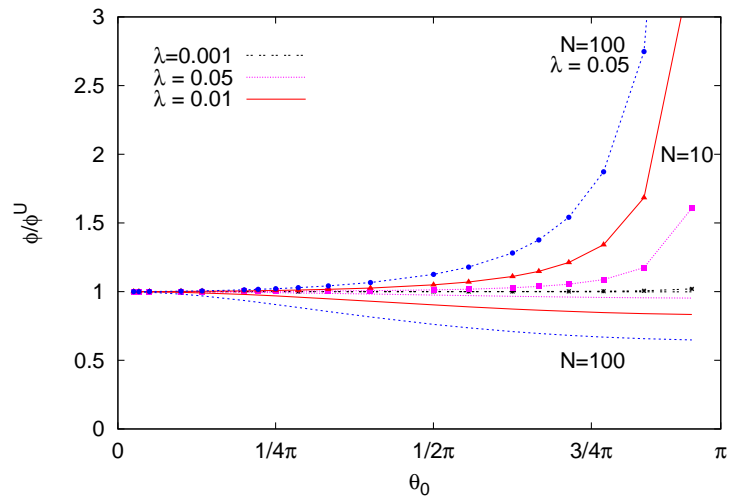}
\caption{Deviation of the exact GP and the perturbative
approximation from the unitary value of the GP
for different initial positions on the Bloch
sphere and for different number of spin environments 
($N=10$ and $N=100$) and different values of $\lambda$.}
\label{fasepertheta0}
\end{figure}

Another feature that can be studied is the deviation from
the unitary value of GP for both different approaches: exact 
and perturbative, as shown in Fig.\ref{fasepertheta0}.
Therein, the GP vs the initial position on the Bloch Sphere
($\theta_0$) is plotted for different environment sizes and
coupling constants. The lines (dotted or solid) correspond
to the exact calculation of the GP while the lines with different
shaped dots are the corresponding perturbative calculations for each 
case. As can be expected, when the coupling between the system and
the environment increases, the perturbative calculation of the
GP differs from the exact one. This difference is more relevant
for larger environments, as can be observed in Fig.\ref{fasepertheta0}
the comparison between an environment comprising $N=10$ spin-1/2 
particles and one with $N=100$ (and the same coupling constant 
$\lambda=0.05$). Not only, is it easy to note that 
this deviation is bigger for larger environments but 
also, it can be observed that
it is more relevant for smaller angles as the number
of environment spins increases.

\begin{figure}[!ht]
\centering
\includegraphics[width=10cm]{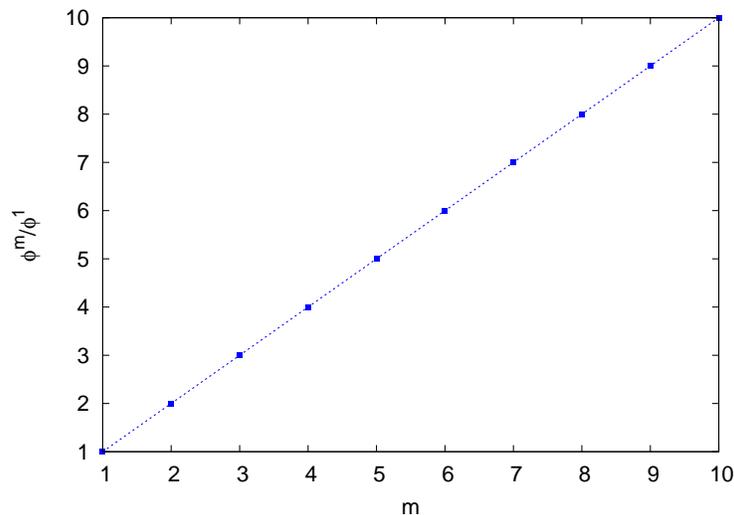}
\caption{ Exact geometric phase of m cycles  
normalized to the exact geometric phase of one cycle
as a function of the number of cycles performed by
the system. The plot results independent of 
$\lambda_{\rm eff}$.}
\label{fasegeometrica2}
\end{figure}

Finally, we can analyze if the phase
retains its geometric nature. Even though we know
that the evolution of the system is not suppused to be
cyclic due to the presence of the environment, 
under certain external conditions it may result a
valid assumption. To that end, we plot
the exact GP of Eq.(\ref{fasegeometrica2}) as a function of
the number of cycles the system evolves normalized by
the exact GP of only one cycle. Therein, we can
note a nice topological feature of the phase:
the phase is dependent on the number of times the
path is traversed, i.e. the winding number. 
In the end, the GP adquired by the system will
be the same. However, this does not mean that 
dephasing will not affect the meassurement of
the GP.  The visibility fringes of any
meassurement made on the system will be reduced due
to the presence of the environment \cite{PRA,Visibility}.

\section{Conclusions}

So far we have studied the corrections to the
geometric phase under a non unitary evolution 
induced by the presence of decoherence. The model is
a further consideration in the spin-spin decoherence
model since we are also considering the self energy
of the spin-1/2 particles comprising the environment.
In addition, it is the simplest approach to study
the nonunitary geometric phase in an Ising chain.

As expected the geometric phase is affected by decoherence.
This correction is quadratic in the coupling constant, so
this affection is negligible when we work in the weak
coupling constant $\lambda/\omega \ll 1$. However, the
correction is linear in the environment's size $N$ as well,
which might have a considerable contribution for 
larger environments.

We have also shown an analytical expression for the environmentally
induced correction to the GP, based in a series expansion in powers of
the system-environment strength. What is worth noticing is that due to
the dependence with the initial angle $\theta_0$, for small angles, there
is not a significant correction induced by the environment, almost all
the geometric character of the phase is in the unitary term. On the contrary,
for a bigger value of $\theta_0$ the corrections to the unitary term become
important. This conclusion is stressed as long as the number of spins in the
environment increases. On the other hand, we have analyzed the validity of
the perturbative calculation of the induced correction to the unitary term,
as it is shown in Fig.\ref{fasepertheta0}. 
Therein, we have shown that the larger environment,
the shorter the values of $\lambda$ for which the perturbative
expansions hold.

Finally, we have studied the geometric nature of the total GP (the one that
includes the environmental correction) changing the single cyclic evolution
ruled by $\tau = 2\pi/\Omega$ to $m$ times $\tau$, with $m$ an
integer number (the winding number). We have shown (see Fig.\ref{fasegeometrica2})
that the total GP can be seen as $m$ times the GP for a single
cyclic evolution, which means that the new phase is geometric because
it depends on the winding number (if you go $m$ times around a path,
the phase is $m$ times the phase for going once around the path).

The main reason why people want to use geometrical
phases for quantum computation is that frequently 
geometrical evolutions are
easier to control and may also be more resistant 
to random noise coming from the environment.
Since experimentally it is much easier to control
the Hamiltonian than the actual state of the system,
the adiabatic evolution is of importance. This means
that the evolution of the system takes a long
time compared to the characteristic dynamical
time scale. This condition must be subtly changed 
in the framework of quantum open systems. The evolution
must take a long
time compared to the characteristic dynamical
time scale but also it must be shorter than the
decoherence time-scale introduced by the presence
of the environment.
In the model we have
studied here, both conditions are fulfilled
since decoherence is not effectively induced
on the system for a realistic small environment.
The results shown herein are particularly 
relevant in the experimental
realizations for the measurements of these phases.

\section{Acknowledgments}

I thank F.C.Lombardo for useful comments.
This work was supported by UBA, CONICET,
and ANPCyT, Argentina.

\end{document}